\documentclass[prb,twocolumn,showpacs,showkeys,amssymb]{revtex4}

\usepackage{graphicx}% Include figure files
\usepackage{dcolumn}% Align table columns on decimal point
\usepackage{bm}% bold math

\begin{document}

%\preprint{}
%\tightenlines

\title{Higher Landau levels contribution \\
 to the energy of interacting electrons in a quantum dot}
\author{Augusto Gonzalez$^{a,b}$, Juan David Serna$^c$,
 Roberto Capote$^d$, and Guillermo Avenda\~no$^a$}
\affiliation{$^a$Departamento de Fisica, Universidad de Antioquia, AA 1040,
 Medellin, Colombia\\
 $^b$Instituto de Cibernetica, Matematica y Fisica, Calle E
  309, Vedado, Ciudad Habana, Cuba\\
 $^c$Department of Physics, University of Arkansas, Fayetteville, Arkansas 72701, USA\\
 $^d$International Atomic Energy Agency, Nuclear Data Section, Vienna A-1400, Austria}

\begin{abstract}
\bigskip
Properly regularized second-order degenerate perturbation theory
is applied to compute the contribution of higher Landau levels to the 
low-energy spectrum of interacting electrons in a disk-shaped quantum dot.
At ``filling factor'' near 1/2, this contribution proves to be larger than 
energy differences between states with different spin polarizations.
After checking convergence of the method in small systems, we show results
for a 12-electron quantum dot, a system which is hardly tractable by means
of exact diagonalization techniques.
\end{abstract}

\pacs{73.21.La, 73.43.Lp}
\keywords {Quantum dots, high magnetic fields, perturbation theory}

\maketitle

\section{Introduction}

The relevant energy scales entering the Hamiltonian of an $N$-electron
system in a quantum dot (qdot) and a magnetic field are the following:
the cyclotronic energy $\hbar\omega_c\sim B$, the dot confinement energy
$\hbar\omega_0$, and the Coulomb characteristic energy $e^2/(\kappa l_B)
\sim \sqrt{B}$. In strong enough fields, the spacing between Landau levels
(LLs), given by $\hbar\omega_c$, is much greater than any other scale, and
one can restrict the Hilbert space to functions built on one-particle 
states from the first LL. This is the 1LL approximation 
\cite{Chakraborty}, which has been widely used to obtain exact solutions
\cite{Laughlin1}, to construct the famous $\nu=1/3$ FQHE functions 
\cite{Laughlin2}, later extended to other filling factors by means of the
Composite Fermion recipe \cite{Jain2} and, in general, has been used to
numerically diagonalize the interacting Hamiltonian \cite{Chakraborty}.

The inclusion of higher LLs in numerical calculations turns out to be
prohibitive, even for relatively ``small'' systems. Consider, for example,
N=12 electrons in a qdot at ``filling factor'' near 1/2, i.e. when the
angular momentum of the electron droplet is $L=-132$. Out of only 78
one-particle states (orbitals) in the 1LL, one can construct 674585 Slater 
determinants, which conform the truncated basis for the 12-particle system
in the 1LL approximation. Taking 78 additional orbitals from each of the
next two LLs causes the basis dimension to be raised to more than 
172 millions, and the diagonalization of the Hamiltonian matrix becomes a
very hard computational task.

In the present paper, we show that a way to circumvent the diagonalization
of these large matrices is the use of properly renormalized degenerate
perturbation theory (PT). We stress that, unlike Monte Carlo and other
methods focusing on the properties of a particular state, by means of
PT we obtain, in a single calculation, an approximation to the energy
spectrum and the corresponding wave functions of the system.

The interest in computing the higher LL contribution to the energies 
relies in the fact that, for intermediate filling factors, this 
contribution may be even larger than energy differences between states
with different spin polarizations \cite{prb}. Thus, a correct
description of spin excitations in a system of interacting electrons
should take account of higher LL effects. Recent work on the issue of
spin excitations in qdots \cite{Haw} has stressed the importance of the
second LL at $\nu\approx 2$, but at lower $\nu$ the higher LL effects are
commonly ignored.

The plan of the paper is as follows. In the next two sections a brief
summary of PT and its regularization by means of Shank extrapolants 
\cite{Shank} and the Principle of Minimal Sensitivity \cite{PMS} is
included for completeness. For simplicity, only spin-polarized systems will
be studied, but any other spin-polarization sector may be treated as well.
Section \ref{results} is devoted to the results. The 2- and 6-electron dots
are used as benchmarks where regularized second-order PT (PT2) is
compared with exact or variational results.  After validation, the method 
is applied to the 12-electron system mentioned above. Concluding remarks 
are given at the end of the paper.

\section{Degenerate perturbation theory}

The 1LL approximation can be seen, from another point of view, as 
first-order degenerate perturbation theory. In fact, writing the 
Hamiltonian in the form: $H=H_0+V$, where $H_0$ describes free 
(spin-polarized) electrons in a magnetic field, and $V=V_{conf}+V_{coul}$ 
accounts for the external confinement and Coulomb interactions, the 
Hamiltonian matrix in the 1LL approximation,

\begin{equation}
H_{ij}^{(1)}=\langle S_i|H|S_j \rangle=E_0 \delta_{ij}+\langle S_i|V|S_j \rangle,
\label{eq1}
\end{equation}

\noindent
where $E_0=N \hbar\omega_c/2$ and $S_i,~S_j$ are Slater determinants
made up from 1LL orbitals, may be seen as the secular matrix
of first-order degenerate perturbation theory \cite{QM}. The degeneracy
subspace is spanned by the $S_i$.

Corrections to (\ref{eq1}) are computed in the standard form \cite{QM}.
The second-order matrix is given by:

\begin{equation}
H_{ij}^{(2)}=E_0 \delta_{ij}+\langle S_i|V+\sum_Z \frac{V|Z\rangle\langle Z|V}
 {E_0-E_0(Z)}|S_j \rangle,
\label{eq2}
\end{equation}

\noindent
where the sum runs over eigenfunctions of $H_0$ in the orthogonal subspace,
$\langle Z|S_i \rangle=0$, and $E_0(Z)=\langle Z|H_0|Z \rangle$.

We will use Eq. (\ref{eq2}) to compute higher LLs contributions to the
energy spectrum of an N-electron qdot. Notice that the dimension of the
secular equation is not increased by the inclusion of the second-order 
corrections. For the largest systems, an energy cutoff, $E_0(Z)-E_0\le
K_{cut}~\hbar\omega_c$, will be imposed to limit the number of states 
entering the sum. We will show results with $K_{cut}=2$, i. e. three
LLs will be included.

\subsection{The orthogonal subspace}

One can explicitly use the fact that $V_{conf}$ and $V_{coul}$ are,
respectively, one- and two-body operators, and exploit their symmetries
(conservation of total angular momentum) in order to carry out the sum only
over intermediate states, $Z$, having nonvanishing matrix elements with
one of the external Slater functions, for example $\langle Z|V|S_j
\rangle\ne 0$.
\vspace{.7cm}

\begin{figure}[ht]
\begin{center}
\includegraphics[width=.8\linewidth,angle=0]{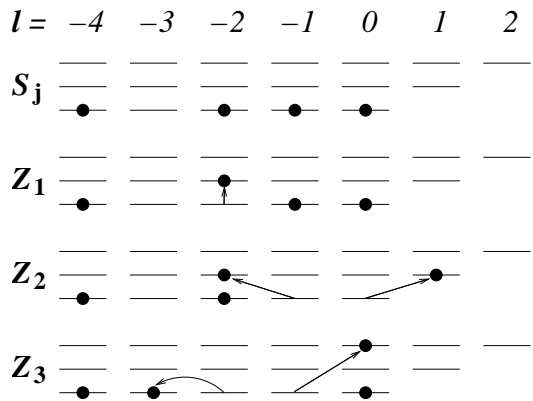}
\caption{\label{fig1} Slater functions in the degeneracy subspace, $S_j$, 
 and in the orthogonal subspace, $Z$, entering the sum in Eq. 
 (\ref{eq2}). The example is for a 4-electron system.}
\end{center}
\end{figure}

In Fig. \ref{fig1}, we have illustrated this statement for the simple
4-electron system. The top of the figure shows the occupation corresponding to
a given $S_j$. Then, the sum will contain functions $Z_1$, where one 
occupied orbital of $S_j$ is raised to an orbital in a higher LL (with the 
same angular momentum, $l$). The sum will also contain functions $Z_2$, 
where two occupied levels of $S_j$ are moved to higher LLs. And, finally, 
functions $Z_3$ in which one occupied level of $S_j$ is moved to an empty 
level in the 1LL and a second occupied orbital is moved to a higher LL
shall also be included. In the first case, both the matrix elements 
of $V_{conf}$ and $V_{coul}$ could be nonzero, whereas in the later two
cases only $V_{coul}$ could have nonvanishing matrix elements. 

\section{Regularization of the perturbative series}

To renormalize the perturbative series (usually an asymptotic series) many
recipes have been invented. In the present paper, we will try 
Shank extrapolants \cite{Shank} and the principle of minimal sensitivity 
(PMS) \cite{PMS}. A variant of the later procedure has been recently applied 
to compute the correlation energy of the Coulomb gas \cite{Sang}.

\subsection{Shank extrapolants}

Shank extrapolants \cite{Shank} are designed to accelerate the 
convergence of numerical series. For any three contiguous values,
$E_i$, $E_{i+1}$ and $E_{i+2}$, we define the extrapolant:

\begin{equation}
F_i=\frac{E_i E_{i+2} - E_{i+1}^ 2}{E_i + E_{i+2} - 2 E_{i+1}}.
\label{extrapolant}
\end{equation}

From the series of extrapolants (which will be called first order) one can 
construct the second-order extrapolants, etc. In our case, we have only 
three values of energy, $E_0$, $E_1$ and $E_2$, obtained from PT0, PT1 and 
PT2, and thus there will be only one extrapolant, $F_0$.

\subsection{The principle of minimal sensitivity}

The PMS starts from the obvious fact that if the Hamiltonian is made to
depend on an artificial trivial parameter $\alpha$: $H\to H(\alpha)$, then
the exact eigenvalues will satisfy the equations:

\begin{equation}
\frac{{\rm d} E}{{\rm d} \alpha}=0.
\label{pms}
\end{equation}

\noindent
The perturbative expansion may not, however, respect these constraints. The
PMS states that an optimal choice for $\alpha$ at a given perturbative order
is the value at which Eq. (\ref{pms}) is satisfied.

In our $N$-electron system, described by the Hamiltonian:

\begin{eqnarray}
H&=&\sum_i \left( \frac{p_i^2}{2 m}+\frac{m\omega_c^2}{8} r_i^2+
 \frac{\hbar\omega_c}{2} l_i\right)+\frac{m\omega_0^2}{2}\sum_i r_i^2
 \nonumber\\
&+& \frac{e^2}{\kappa}\sum_{i<j}\frac{1}{r_{ij}},
\end{eqnarray}

\noindent
the introduction of trivial terms and a scaling of coordinates lead to:

\begin{eqnarray}
H&=&\frac{\hbar\omega_c'}{2} \sum_i \left( \frac{p_i^2}{2}+\frac{r_i^2}{2}
 +l_i\right) \nonumber\\
 &+&\frac{\hbar}{\omega_c'}\left(\omega_0^2+\frac{\omega_c^2}{4} 
 -\frac{\omega_c'^2}{4}\right) \sum_i r_i^2\nonumber\\
 &+&\left( \frac{\hbar\omega_c}{2}-\frac{\hbar\omega_c'}{2}\right) L
 +\frac{e^2}{\kappa}\sqrt{\frac{m\omega_c'}{2\hbar}}\sum_{i<j}\frac{1}{r_{ij}},
\label{hamilt}
\end{eqnarray}

\noindent
where $\hbar\omega_c'$ is an artificial parameter (magnetic field) and
$L=\sum_i l_i$. The first sum in Eq. (\ref{hamilt}) will be taken as the
unperturbed Hamiltonian, $H_0$, and the subsequent terms as the 
perturbation, $V$. Notice that the term 
$(\hbar\omega_c/2-\hbar\omega_c'/2) L$ in $V$ will give no 
contribution to the second-order correction because $[H_0,L]=0$.

The parameter $\hbar\omega_c'$ will be fixed from the condition 
(\ref{pms}). Working whithin PT1, one can interpret $\hbar\omega_c'$ as 
a variational parameter. In PT2, however, the point in which (\ref{pms}) 
is satisfied in our examples is a local maximum, as will be seen below.

\section{Results}
\label{results}

The qdot parameters to be used are borrowed from Ref. \onlinecite{prb}. 
GaAs effective mass, $m=0.067 m_0$, and relative dielectric constant
$\kappa=12.5$ are employed.
The confinement potential is parabolic with $\hbar\omega_0=3$ meV. The
bare Coulomb interaction is weakened by a factor 0.8 to
approximately account for quasi-bidimensionality (instead of exact 
bidimensionality). We will only deal with spin-polarized systems, thus
the Zeeman energy is a trivial constant and will not be included.

\subsection{Two electrons}

In the two-electron dot, calculations may be carried out semi-analytically.
The results for the ground-state energy of the triplet state with $L=-1$
at $B=8$ T are the following: $E_0=13.82$, $E_1=20.28$, and $E_2=20.16$ meV. 
The exact energy is $E=20.17$ meV. Notice that the higher LL contribution 
to $E$ is -0.12 meV, and that $E_2$ is very close to the exact value.

Regularization of the perturbative series by means of the Shank 
extrapolant yields a value $F_0$ which practically coincides with $E_2$. 
The difference lies in the fifth significant figure, not written above.

\begin{figure}[ht]
\begin{center}
\includegraphics[width=.65\linewidth,angle=-90]{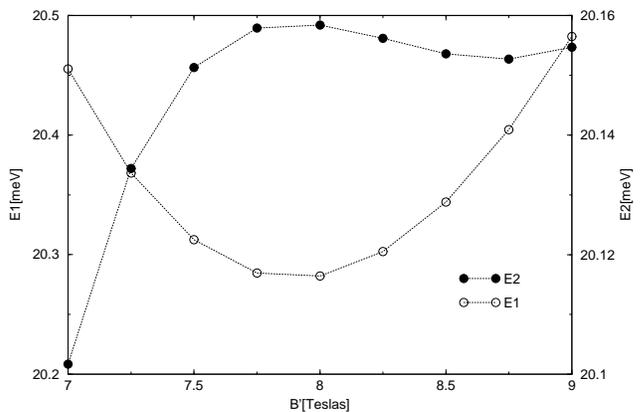}
\caption{\label{fig2} $E_1$ and $E_2$ as a function of $B'$ for the 
 two-electron system at $B=8$ T.}
\end{center}
\end{figure}

On the other hand, as a function of the PMS parameter, $B'$, we obtain the
curves shown in Fig. \ref{fig2} for the magnitudes $E_1(B')$ and 
$E_2(B')$.

Note that the position of the minimum in $E_1(B')$ practically coincides 
with the position of the maximum in $E_2(B')$. These are the physically 
acceptable values of $B'$ according to (\ref{pms}). Note also that the 
PMS-regularized ground state energy has the same significant figures as 
the unregularized second-order result, $E_2(B)$.

\subsection{Six electrons}

The exact diagonalization (variational) results of Ref. \onlinecite{prb}
show that at $B=8$ T (filling factor near 1/2) the contribution of the
second and third LLs to the energy eigenvalues is around -0.4 meV, i. e.
around -0.06 meV per electron, as in the two electron system. This
magnitude is greater than the energy differences between states with
different spin polarizations.

\begin{figure}[ht]
\begin{center}
\includegraphics[width=.65\linewidth,angle=-90]{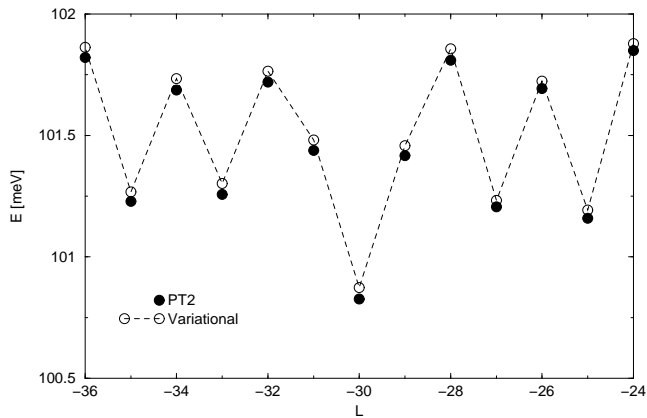}
\caption{\label{fig3} Variational and PT2 results for the
 low-lying states in the 6-electron dot. $B=8$ T.}
\end{center}
\vspace{.3cm}
\end{figure}

The lowest spin-polarized states in each angular momentum
tower (the yrast spectrum) are shown in Fig. \ref{fig3}. Energy jumps
between adjacent angular momentum states are about 0.6 meV. The PT2
results are shown to lay 0.04 meV below the variational ones, that is,
almost inside the estimated error bars which are about 0.03 meV. These
PT2 results are obtained with $K_{cut}=2$, as mentioned above.

Regularization of the perturbative series leads to results very similar
to those in the two-electron dot: the significant figures (in this case
five) are not changed. This means that one can in practice estimate
the ground state energy from the PT2 result.

\begin{figure}[ht]
\includegraphics[width=0.95\linewidth,angle=0]{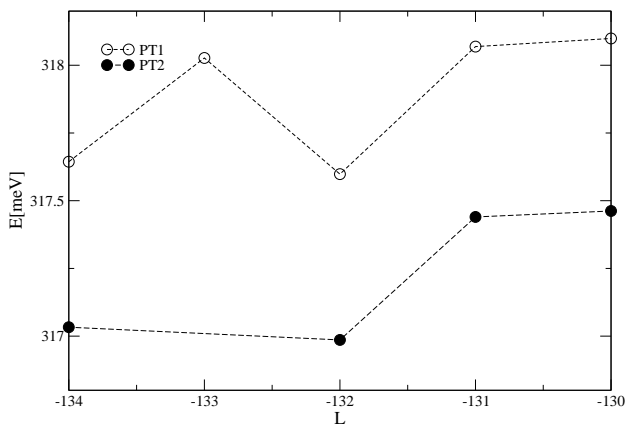}
\caption{\label{fig4} PT1 and PT2 estimates of the low-lying spin-polarized
 energy levels of twelve electrons. $B=10$ T.}
\end{figure}

\subsection{Twelve electrons}

We show in Fig. \ref{fig4} the perturbative low-lying spectrum for the
12-electron dot. In this situation, the perturbative results are the only ones at our
disposal because exact diagonalization is practically impossible. The
magnetic field is 10 T in this figure. Notice the energy jumps between
adjacent angular momentum states, as in the 6-electron dot. The contribution
of the higher LLs to the energy eigenvalues is around -0.6 meV, that is,
-0.05 meV per electron.

Only the lowest eigenvalues in each angular momentum tower are drawn in Fig.
\ref{fig4}, but the next 20-30 eigenvalues are reliable as well. The generation
of the second-order matrix, Eq. (\ref{eq2}), for a single $L$ value takes
around one week CPU time in a small computer cluster with ten processors at
2.4 GHz. The PT2 matrix is a factor of ten less sparse than the PT1 matrix,
occupying around 15 GB of hard disk. Consequently, diagonalization by means
of a Lanczos algorithm takes a factor of ten more time.

\section{Concluding remarks}

We have shown the feasibility of computing the energy eigenvalues and eigenfunctions
of relatively large electronic quantum dots in a magnetic field by using second-order
degenerate perturbation theory to take account of the contribution of the higher
LLs. This contribution proves to be of the same order of energy differences between
states with different spin polarizations. The constructed effective Hamiltonian
in the 1LL can be diagonalized by means of standard Lanczos techniques. We
presented calculations for the spin-polarized states of a 12-electron quantum dot at
filling factor near 1/2. When three LLs are included, the dimension of the truncated
Hilbert space for this problem is larger than 170 millions.

From a more general point of view, a similar scheme could be applied to any problem
with well differentiated energy scales, in which one could identify, at least
approximately, degeneracy subspaces. Examples of such problems are the
study of mixing between vibrational and electronic degrees of freedom in
relatively large molecules, or the study of mixing between intra- and inter-shell
excitations in relatively large nuclei.

\begin{acknowledgments}
The authors acknowledge the Committee for Research of the
Universidad de Antioquia (CODI) and the Colombian Institute for Science and
Technology (COLCIENCIAS) for support.
\end{acknowledgments}

\end{document}